\documentclass[a4paper,11pt]{article}
\pdfoutput=1 % if your are submitting a pdflatex (i.e. if you have
\usepackage{jheppub}
\usepackage[T1]{fontenc}
\usepackage{graphicx}
\usepackage{color}
\usepackage{hyperref}
\usepackage{slashed}
\usepackage{subfigure}

\title{\boldmath  Emergent Lorentz symmetry near fermionic quantum critical points in two and three dimensions}

\author[a,1]{Bitan Roy, \note{Corresponding author.}}
\author[b,c]{Vladimir Juri\v ci\' c,}
\author[d]{Igor F. Herbut}

\affiliation[a]{Condensed Matter Theory Center, Department of Physics,
University of Maryland, College Park, MD 20742, USA}
\affiliation[b]{Nordita,  Center for Quantum Materials,  KTH Royal Institute of Technology and Stockholm University, Roslagstullsbacken 23,  10691 Stockholm,  Sweden}
\affiliation[c]{Institute for Theoretical Physics, Center for Extreme Matter and Emergent Phenomena,
Utrecht University, Leuvenlaan 4, 3584 CE Utrecht, The Netherlands}
\affiliation[d]{Department of Physics, Simon Fraser University, Burnaby, British Columbia V5A 1S6, Canada}

\emailAdd{broy@umd.edu}
\emailAdd{juricic@nordita.org}
\emailAdd{iherbut@sfu.ca}

\abstract{We study the renormalization group flow of the velocities in the field theory describing the coupling of the massless quasi-relativistic fermions to the bosons through the Yukawa coupling, as well as with both bosons and fermions coupled to a fluctuating $U(1)$ gauge field in two and three spatial dimensions. Different versions of this theory describe quantum critical behavior of interacting Dirac fermions in various condensed-matter systems. We perform an analysis using one-loop $\epsilon-$expansion about three spatial dimensions, which is the upper critical dimension in the problem. In two dimensions, we find  that velocities of both charged fermions and bosons ultimately flow to the velocity of light, independently of the initial conditions, the number of fermionic and bosonic flavors, and the value of the couplings at the critical point. In three dimensions, due to the analyticity of  the gauge field propagator, both the $U(1)$ charge and the velocity of light flow, which leads to a richer behavior than in two dimensions. We show that all three velocities ultimately flow to a common terminal velocity, which is non-universal and different from the original velocity of light. Therefore, emergence of the Lorentz symmetry in the ultimate infrared regime seems to be a rather universal feature of this class of theories in both two and three dimensions.}


\begin{document}
\maketitle
\flushbottom

\section{Introduction}

  Lorentz invariance appears to be a fundamental symmetry of nature as we know it\cite{kostelecky}. Nevertheless, there have been several attempts to understand it as a property that emerges only at low energies~\cite{horava, nielsen1, nielsen2, anber, bednik, sibiryakov, kharuk}. That this is indeed possible in principle has been known for quite a while in condensed matter physics, where Lorentz symmetry often appears in symmetry-poor lattice models near the fixed points of the renormalization group flow~\cite{vafek, djlee, HJR}.

  One such example is the quantum critical point of the Gross-Neveu-Yukawa model of Dirac fermions coupled to bosonic order parameter via a Yukawa term, which  depending on the precise pattern of symmetry breaking, is believed to describe a variety of zero temperature phase transitions in Dirac-like condensed matter systems, such as monolayer graphene and twisted bilayer graphene in two dimensions, or Weyl semimetals in three spatial dimensions~\cite{HJV, RJH, Lee, roy-yang, yao, nandkishore, roy-dassarma}. A small difference in the velocities of bosons and fermions in such a theory is an irrelevant perturbation at the critical point. Another example is the flow of Fermi velocity towards the fixed velocity of light in graphene, under the influence of gauge interaction~\cite{vozmediano}. In this case the velocity of the gauge field cannot renormalize and stays fixed under the scale transformation, due to the fact that matter is confined to a space of dimension lower than that in which the gauge field lives. In 3+1 dimensions, Dirac fermions coupled to the gauge field also ultimately acquire Lorentz symmetry with the terminal common velocity towards which both the initial velocities of fermions and gauge field scale \cite{isobe}. The scale dependence of the velocity of light stems from the analyticity of the gauge field propagator in 3+1 dimensions.

Emergence of the Lorentz symmetry at low energies in a boson-fermion coupled theory has been studied in Ref.~\cite{anber}, but only in 3+1 space-time dimensions and assuming the bosons to be gauge-neutral objects. Moreover, a similar problem has been analyzed in the context of the purely bosonic~\cite{iengo} and boson-fermion coupled~\cite{gomes} Lifshitz theories and Lorentz-symmetry breaking quantum electrodynamics~\cite{kostelecky}.

Motivated by these examples, in this paper we study a more general model of Dirac fermions coupled to O(2) bosonic field via Yukawa term, and with fermions and bosons, both or separately, additionally coupled to a gauge field, both in 3+1 and lower space-time dimensions. We assume arbitrary initial velocities of all three fields, and monitor their change under the scaling transformation. We find that ultimately in the deep infrared regime the Lorentz symmetry with one common velocity is {\it always} established, although at intermediate scales the flow is rather non-universal, and may even be non-monotonic. This is a demonstration that in a rather broad class of Lorentz-violating theories with fermions, charged or charge neutral bosons and gauge fields, different couplings between these separate sectors yield a universal outcome at low-energies: the Lorentz invariance with a common velocity established by the interaction.

  This paper is organized as follows. In Sec.~\ref{GNY}, we introduce the Gross-Neveu-Yukawa theory and set up the formalism. In Sec.~\ref{RG-equation}, we derive the renormalization group (RG) flow of the velocity for the matter and gauge fields in both (2+1) and (3+1) dimensions within the framework of one-loop $\epsilon$-expansion. In Sec.~\ref{RG-2D}, we analyze these RG flows in (2+1) dimensions, while in Sec.~\ref{RG-3D} a similar analysis has been performed in 3+1 dimensions. As we will show, in both cases the flow of the velocity in the infrared reaches a terminal velocity, and therefore Lorentz symmetry emerges at low energies. We summarize the results and discuss related topics in Sec.~\ref{Conclusion}. Details of the calculations of the self-energy for matter and gauge fields are presented in Appendices.

\section{Gross-Neveu-Yukawa theory}\label{GNY}

We consider the  system of quasi-relativistic fermions coupled to a fluctuating $U(1)$ gauge field, as well as to a bosonic order parameter field through the Yukawa coupling ($g$). The complex bosonic order parameter field displays the $O(2)$ symmetry and carries $N_B$ flavors. The four-component massless Dirac fermions are enriched by $N_F$ copies, and interact with the gauge field through charge $e$. In contrast, the bosonic field is assumed to be a composite object of two fermionic fields and thus carries a charge $2e$. The dynamics of the system is then described by the following imaginary time (Euclidean) action~\cite{RJH}
\begin{equation}\label{action}
S=\int d^{D}x \; d\tau \: \: \left( L_F + L_B +L_{BF}+ L_{EM} \right),
\end{equation}
where
\begin{eqnarray}
L_F &=& \bar{\Psi} \left[ \gamma_0 (\partial_0 - i e \gamma_5 A_0) + \gamma_j v_F (\partial_j - i \frac{e}{c} \gamma_5 A_j) \right] \Psi , \nonumber \\
L_B &=& |(\partial_0 + 2 i e A_0) \Phi|^2 + v^2_B |(\partial_j + 2 i \frac{e}{c} A_j) \Phi|^2 + m^2 |\Phi|^2 +\frac{\lambda}{2} |\Phi|^4, \nonumber \\
L_{BF}&=& g \left[ \left( \mbox{Re}\Phi \right) \bar{\Psi} \Psi +\left( \mbox{Im}\Phi \right) \bar{\Psi} i \gamma_5 \Psi \right], \nonumber \\
L_{EM} &=& \frac{1}{4} F_{\mu \nu} F_{\mu \nu}.
\end{eqnarray}
Here, $\bar{\Psi} \equiv \bar{\Psi}(\vec{x}, \tau)$, $\Psi \equiv \Psi(\vec{x}, \tau)$, $\Phi \equiv \Phi(\vec{x}, \tau)$, and $\bar{\Psi}=\Psi^\dagger \gamma_0$ is an independent Grassmann variable. The fermionic and bosonic velocities are represented by $v_F$ and $v_B$, respectively, and $F_{\mu \nu}$ is the electromagnetic field strength tensor. The velocity of light is denoted by $c$. The $\gamma$ matrices satisfy the standard anticommutation Clifford algebra $\left\{ \gamma_\mu, \gamma_\nu \right\}=2 \delta_{\mu \nu}$ for $\mu, \nu=0, 1, \cdots, D$, and $\left\{ \gamma_\mu, \gamma_5 \right\} =0$. Summation over repeated indices is assumed. The explicit form of $L_{EM}$ in terms of temporal ($A_0$) and spatial components ($A_j$) of the gauge fields in real (Minkowski) time and in $3+1$ dimensions is given by Eq.~(\ref{em-realtime}).

The above field theory can emerge as an effective description near the quantum phase (continuous) transitions of massless Dirac fermions towards the formation of an ordered phase. For instance, if we neglect the coupling between the complex bosonic and the fluctuating gauge fields, the above effective action describes a quantum phase transition out of Dirac semimetal into an $O(2)$ symmetry breaking insulating phase. The physical meaning of the ordered phase depends on the representation of the spinor field. With the appropriate definition, the above theory can describe the transition into the Kekule valence bond insulator in graphene~\cite{RJH} and surface (top and bottom)-hybridizing and time-reversal symmetric insulator in thin topological insulators~\cite{seradjeh}. When a slightly modified spinor representation is chosen and the bosonic field is charged, the above field theory captures the universality class of a phase transition from the Dirac semimetal to a spin-singlet $s$-wave superconductor in graphene and surface states of topological insulators~\cite{RJH}. If we neglect the boson-gauge field coupling, the ordered phase in three dimensional Dirac or Weyl semimetal represents an axionic insulator, while the complete action in Eq.~(\ref{action}) is pertinent near the transition to an s-wave superconductor~\cite{roy-dassarma, ohsaku}. When the bosonic field is assumed to be gauge-neutral, the theory becomes the celebrated Nambu-Jona-Lasinio model for spontaneous chiral symmetry breaking in 3+1 dimensions~\cite{NJL}. With these motivations in mind, we will  focus here on the flow of three velocities, $v_F$, $v_B$ and $c$ in two and three spatial dimensions, that is in 2+1 and 3+1 space-time dimensions, and establish the emergence of the Lorentz invariance as the three velocities approach a common terminal velocity at low energies.

\begin{figure*}[htb]
\begin{center}
\includegraphics[width=15.00cm,height=6.50cm]{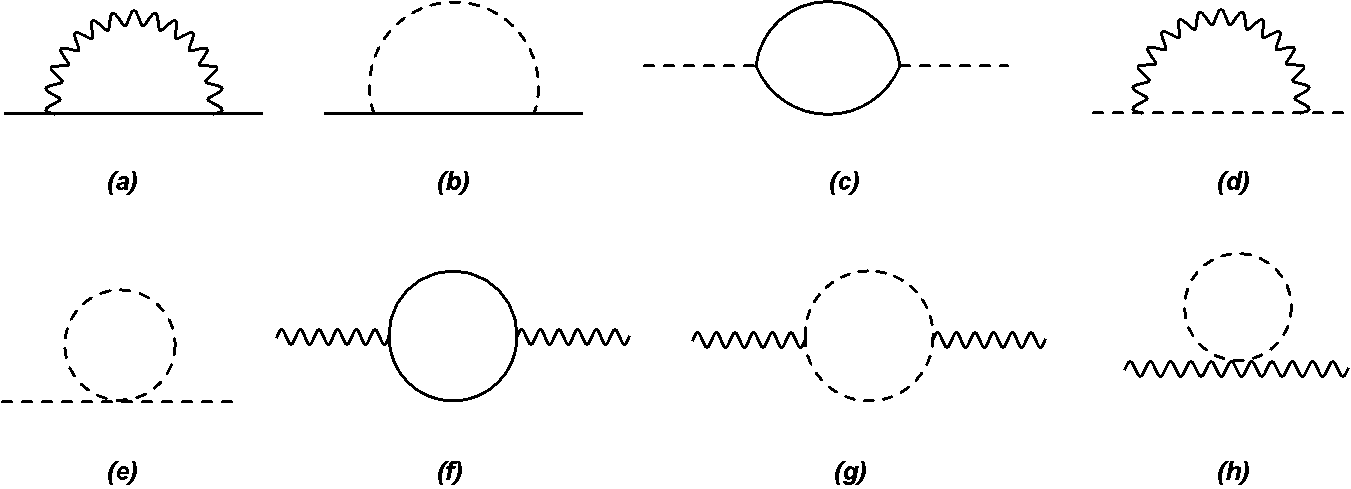}
\end{center}
\caption[] {Self-energy diagrams for Dirac fermion [(a) and (b)], bosonic [(c), (d) and (e)] and electro-magnetic gauge [(f), (g) and (h)] fields. Here solid, dashed and wavy lines represent fermionic, bosonic and gauge fields, respectively. Diagram $(e)$ only renormalizes the bosonic mass ($m$). Notice that diagrams (f), (g) and (h) are pertinent only in three spatial or 3+1 space-time dimensions. }\label{self-energy-diag}
\end{figure*}

Notice that all coupling constants in the action, namely $e, g$, and $\lambda$ are \emph{marginal} in three spatial dimensions ($D=3$). Therefore, we perform a perturbative $\epsilon$-expansion around  \emph{the upper critical dimension}, where $\epsilon=3-D$ to capture the emergent low-energy phenomena. The fermionic, bosonic and gauge field propagators read, respectively,
\begin{eqnarray}
G_F(\omega, \vec{k}) &=& \frac{i \left( \gamma_0 \omega + v_F \gamma_j k_j \right) }{\omega^2+ v^2_F k^2}, \\
G_B(\omega, \vec{k}) &=& \frac{1}{\omega^2 +v^2_B k^2 +m^2}, \\
D_{\mu \nu}(\omega, \vec{k}) &=& \frac{\delta_{\mu \nu}}{\varepsilon \left( \omega^2 + c^2 k^2 \right)^{\frac{D-1}{2}}} ,
\end{eqnarray}
where $\varepsilon$ is the permittivity of the medium. The velocity of light in the medium is $c=1/\sqrt{\varepsilon \mu}$, where $\mu$ is the permeability of the medium.

 The structure of the gauge field propagator $D_{\mu \nu}(\omega, \vec{k})$ is such that its inverse is the usual, analytic, quadratic function of the four-momentum in $D=3$ (spatial dimensions). However, when $D<3$, $D^{-1} _{\mu \nu}(\omega, \vec{k})$ becomes manifestly {\it non-analytic}. In particular, in two spatial dimensions the gauge field propagator is defined here as  $D_{\mu \nu}(\omega, \vec{k}_\perp)=1/\left(\varepsilon \sqrt{\omega^2+ c^2 k^2_\perp}\right)$, with $\vec{k}_\perp=(k_1,k_2)$. This non-analyticity arises from the ``projection" of the gauge-field, which lives in 3+1 dimensions, onto the $2+1$-dimensional ``brane"  to which the matter fields are confined \cite{gorbar, mastropietro}. This difference in the analyticity of the gauge field propagator between $D=3$ and $D<3$ dimensions will prove to be important for the behavior of the coupling constants under renormalization~\cite{z}.

We here also choose to work with the Feynman gauge for the sake of simplicity, since the universal behavior of the theory should not depend on the choice of gauge. As the theory does not possesses the Lorentz symmetry at the bare level, we also carefully maintain the distinction between space and time co-ordinates, and show that only inside the deep infrared regime is the full space-time (Lorentz) symmetry achieved.

\section{Renormalization group flow of velocities}\label{RG-equation}

 Using the $\epsilon$-expansion as a tool for studying the quantum-critical behavior in the infrared regime, we obtain the renormalized action $S_R=\int d^D x d\tau L_{R}$, where
\begin{eqnarray}
L_{R} &=& \bar{\Psi} \left[ Z_\Psi \gamma_0 \left( \partial_0 -i e \gamma_5 A_0 \right) + Z_\Psi Z_{v_F} \gamma_j \left( \partial_j-i \frac{e}{c} A_j \right) \right] \Psi + Z_\Phi |\left(\partial_0+ 2 i e A_0 \right)\Phi|^2  \nonumber \\
&+& Z_\Phi Z^2_{v_B} v^2_B |\left(\partial_j + 2 i \frac{e}{c} A_j \right)\Phi|^2
+ Z_m m^2 |\Phi|^2 + Z_\lambda \frac{\lambda}{2} |\Phi|^4 + Z_g L_{B-F} + Z_A L_{EM},
\end{eqnarray}
and $Z_j$'s are renormalization constants. In dimensions $D=3-\epsilon$, for all $\epsilon>0$,  $Z_A=1$ exactly. This is a direct consequence of the non-analyticity of the gauge field propagator below three spatial dimensions, as discussed above: integrating out degrees of freedom near the  UV cutoff cannot produce terms that would be non-analytic in fields or momenta in the remaining low-energy action~\cite{z}. In other words, neither the charge ($e$) nor the velocity of light ($c$) receives any perturbative corrections for $\epsilon>0$. The computation of the self-energy diagrams for Dirac fermion and bosonic field, shown in Fig.~\ref{self-energy-diag}(a)- \ref{self-energy-diag}(e), yields the following renormalization factors
\begin{eqnarray}
Z_\Psi &=& 1 + \frac{e^2}{4 \pi^2 \varepsilon} \: \left(1-3\;  \frac{v^2_F}{c^2}\right) \frac{1}{c (c+v_F)^2} \; \frac{1}{\epsilon}-\frac{g^2}{4 \pi^2} \: \frac{1}{v_B (v_F + v_B)^2} \; \frac{1}{\epsilon}, \label{field-fermion}\\
Z_{v_F} &=& Z^{-1}_\Psi \left[ 1- \frac{e^2}{12 \pi^2 \varepsilon}  \left( 1+\frac{v^2_F}{c^2} \right) \frac{ 2 c +v_F}{(c+v_F)^2} \; \frac{1}{\epsilon}
-\frac{g^2}{12 \pi^2} \; \frac{v_F + 2 v_B}{v_F v_B (v_F+v_B)^2 } \; \frac{1}{\epsilon} \right], \label{velocity-fermion}\\
Z_\Phi &=& 1- \frac{g^2}{8 \pi^2} \; \frac{N_F}{v^3_F} \; \frac{1}{\epsilon} -\frac{e^2}{\pi^2 \varepsilon} \left( \frac{c^2+2 c v_B -v^2_B}{c^3 v_B (c+v_B)} \right) \; \frac{1}{\epsilon}, \label{field-boson} \\
Z_{v_B} &=& Z^{-1/2}_\Phi \left[1-\frac{g^2}{16 \pi^2} \; \frac{N_F}{v_F v^2_B} \frac{1}{\epsilon} - \frac{e^2}{2 \pi^2 \varepsilon} \left(  \frac{38 c^3 +31 c^2 v_B -6 c v^2_B -3 v^3_B}{15 c^3 v_B (c+v_B)^2} \right) \; \frac{1}{\epsilon} \right]. \label{velocity-boson}
\end{eqnarray}
We here used the minimal subtraction scheme, and the divergent part of each diagram is evaluated by performing the momentum integrals using dimensional regularization, which preserves the gauge invariance. Details of the calculation of fermionic and bosonic self-energies are presented in the Appendix~\ref{append-fermion-SE} and \ref{append-boson-SE}, respectively. As well known, in this formalism the poles $1/\epsilon$ capture the logarithmically divergent contributions.

\begin{figure}[htbp]
\centering
\subfigure[]{
\includegraphics[scale=0.55]{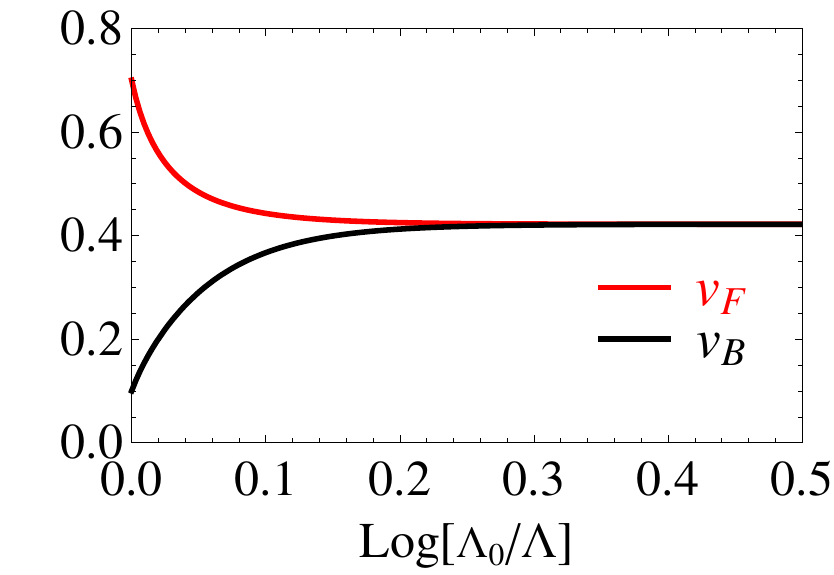}
\label{flow-decoupled-a}
}
\subfigure[]{
\includegraphics[scale=0.55]{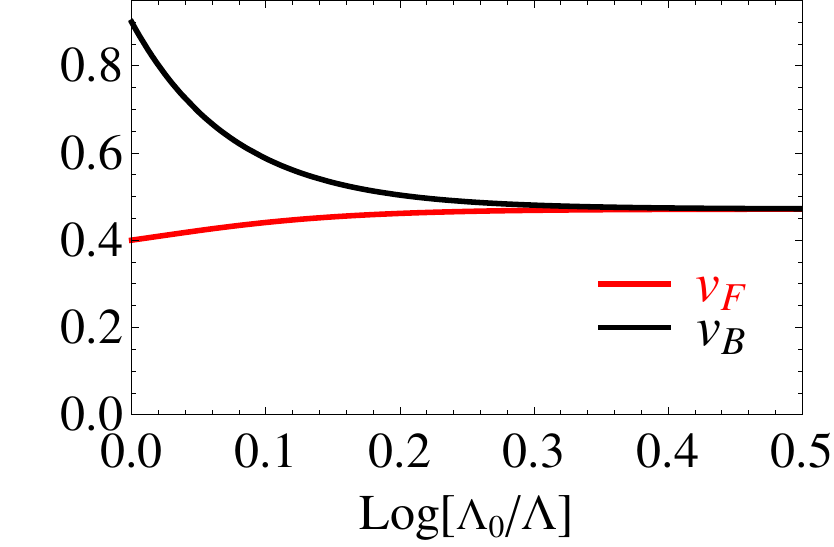}
\label{flow-decoupled-b}
}
\subfigure[]{
\includegraphics[scale=0.55]{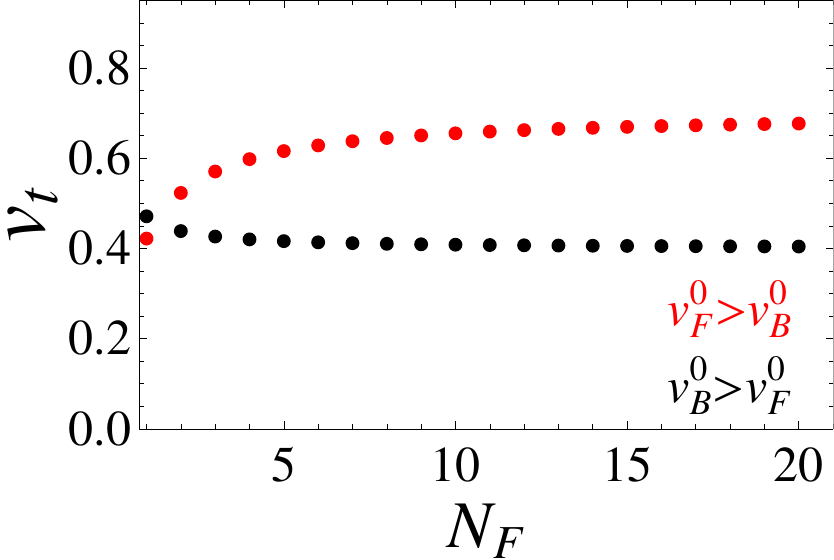}
\label{flow-decoupled-c}
}
\caption[] { Renormalization group flows of Fermi velocity (red curves) and bosonic velocity (black curves) when (a) $v^0_F>v^0_B$, and (b) $v^0_B>v^0_F$, for $N_F=1$. (c) Dependence of the terminal velocity ($v_t$) on the number of fermion flavors ($N_F$) for $v^0_F=0.7$ and $v^0_B=0.1$ (red dots), and $v^0_F=0.4$, $v^0_B=0.9$ (black dots). Throughout we set the Yukawa coupling $g=1$. Here, $\log \left[\Lambda_0/\Lambda \right]$ is the renormalization group time, where $\Lambda_0$ is the ultraviolet cutoff, and $\Lambda$ is the running infrared cut-off.}\label{flow-decoupled}
\end{figure}

From the above renormalization factors, we obtain the following flow equations in the infrared regime for the Fermi velocity ($v_F$) and the bosonic velocity ($v_B$)
\begin{eqnarray}\label{fermion-flow}
\beta_{v_F} = \frac{4 \; \alpha_F \; v_F}{3 (1+ \frac{v_F}{c})^2}  \left[ 1+2 \left( \frac{v_F}{c} \right) + \left(\frac{v_F}{c} \right)^2-4 \left( \frac{v_F}{c} \right)^3 \right]-\frac{4 \; g^2}{3} \frac{v_F}{v_B (v_F+v_B)^2} \left[ 1-\frac{v_B}{v_F}\right], \nonumber \\
\end{eqnarray}
\begin{eqnarray}\label{boson-flow}
\beta_{v_B} = \frac{4 \; \alpha_F \; v_F}{15 (1+\frac{v_B}{c})^2} \left[ 23-14 \left( \frac{v_B}{c} \right)-21 \left( \frac{v_B}{c} \right)^2 +12 \left( \frac{v_B}{c} \right)^3 \right] - \frac{N_F}{2} g^2 \frac{v_B}{v^3_F} \left[ 1-\left( \frac{v_F}{v_B} \right)^2 \right], \nonumber \\
\end{eqnarray}
after taking $e^2/(8\pi^2) \to e^2$, $g^2/(8\pi^2) \to g^2$, and $\alpha_F=e^2/(\varepsilon c^2 v_F)$ can be considered as the fine structure constant in the medium. We are assuming that the Yukawa coupling $g$ is at its fixed point value, $g^2 \sim \epsilon$. In the following sections, we show that terminal velocity of fermions and bosons is insensitive of the Yukawa coupling ($g$). Notice that flow equations are independent of the bosonic vertex coupling $\lambda$ [see Eq.~(\ref{action})].

Although in two spatial dimensions the velocity of light in the medium does not receive any perturbative corrections, the situation is dramatically different in three dimensions. In $D=3$ the gauge field propagator is an analytic function of the four-momentum. Consequently both permittivity ($\varepsilon$) and permeability ($\mu$) of the medium get renormalized [from diagrams in Fig.~\ref{self-energy-diag} $(f)$-$(h)$] due to the coupling of fluctuating gauge field with massless Dirac fermions and the bosonic field. Detailed analysis of polarization bubbles is shown in Appendix~\ref{gauge-SE-appendix}. Respectively, the flow of the fine structure constant ($\alpha_F$) and the velocity of light ($c$) in the medium reads as
\begin{eqnarray}
\beta_{\alpha_F} &=&- \frac{4}{3}\; \alpha_F^2 \; \left[N_F + N_B \; \frac{c^3 v_F}{v^4_B} \right] \; \delta_{D,3},\label{alpha-flow} \\
\beta_{c}&=&-\frac{2}{3} \; \alpha_F\;  c \; \left[ N_F \left(1-\frac{v^2_F}{c^2} \right) +N_B \; \frac{c v_F}{v^2_B} \left( \frac{c^2}{v^2_B}-1\right) \right] \; \delta_{D,3}, \label{c-flow}
\end{eqnarray}
where ``$\delta$" is the Kronecker delta function. Here flow of the electric charge ($e$) is captured by Eq.~(\ref{alpha-flow}).

\section{Terminal velocity in two and three dimensions}\label{Rg-analysis}

We now proceed with the analysis of the flow equations for $v_F$, $v_B$ and $c$ in two and three spatial dimensions separately.

\subsection{Renormalization group flow of the velocities in two dimensions}\label{RG-2D}

Let us first focus on two dimensions, where neither the velocity of light nor the $U(1)$ charge receives any perturbative corrections.

  For simplicity, we now neglect the coupling of Dirac fermions and bosonic field with the fluctuating gauge field. Such situation can be addressed by setting $e=0$ in the flow equations (\ref{fermion-flow}) and (\ref{boson-flow}). We numerically analyze the RG flow equations for $v_F$ and $v_B$, and the results are shown in Fig.~\ref{flow-decoupled}, for $v^0_F>v^0_B$ [Fig.~\ref{flow-decoupled-a}] and $v^0_B>v^0_F$ [Fig.~\ref{flow-decoupled-b}]. The quantities with superscript ``$0$'' denote the bare values. If the bare velocity of fermions is greater than that of bosons ($v^0_F>v^0_B$) during the RG flow  these two velocities approach each other, and at an infrared stable fixed point both fermionic and bosonic degrees of freedom acquire a unique terminal velocity $v_F=v_B\equiv v_t$, where $v^0_B< v_t< v^0_F$, as shown in Fig.~\ref{flow-decoupled-a}~\cite{Lee, nandkishore}. If, on the other hand, the bare velocity of boson is larger than that of fermions ($v^0_B> v^0_F$), ultimately these two velocities approach an infrared stable terminal velocity $v_t$, where $v^0_F< v_t< v^0_B$, as shown in Fig.~\ref{flow-decoupled-b}. The dependence of the terminal velocity on initial velocities ($v^0_F, v^0_B$) may not be too surprising. For fixed $v^0_B$, the terminal velocity increases monotonically with increasing $v^0_F$, irrespective of whether $v^0_F> v^0_B$ or $v^0_F< v^0_B$. Furthermore, the terminal velocity is insensitive to the strength of the Yukawa coupling ($g$), although the strength of $g$ controls the intermediate speed of these two species. Therefore, for a fixed number of fermionic and bosonic species, $N_F$ and $N_B$, the velocity acquires a universal value independent of the value of the Yukawa coupling at the quantum-critical point governing the transition into a symmetry-broken state.

The dependence of the terminal velocity $v_t$ on fermionic ($N_F$) and bosonic ($N_B$) flavor number is more interesting. Notice that flow equations for $v_F$ and $v_B$ do not depend on $N_B$, and consequently $v_t$ is independent of the number of bosonic flavors. However, the terminal velocity $v_t \to v^0_F$ as $N_F \to \infty$, irrespective of whether $v^0_F>v^0_B$ or $v^0_F<v^0_B$, as shown in Fig.~\ref{flow-decoupled-c}. Such behavior can be appreciated from the appearance of $N_F$ in the second term in Eq.~(\ref{boson-flow}), which dictates that with  increasing number of fermionic flavors $N_F$, boson velocity acquires a larger boost (either increasing or decreasing depending on whether $v^0_F>v^0_B$ or $v^0_F<v^0_B$, respectively) and the terminal velocity asymptotically approaches the bare fermionic velocity ($v^0_F$) for a large number of fermionic species ($N_F \to \infty$).

\begin{figure}[htbp]
\centering
\subfigure[]{
\includegraphics[scale=0.55]{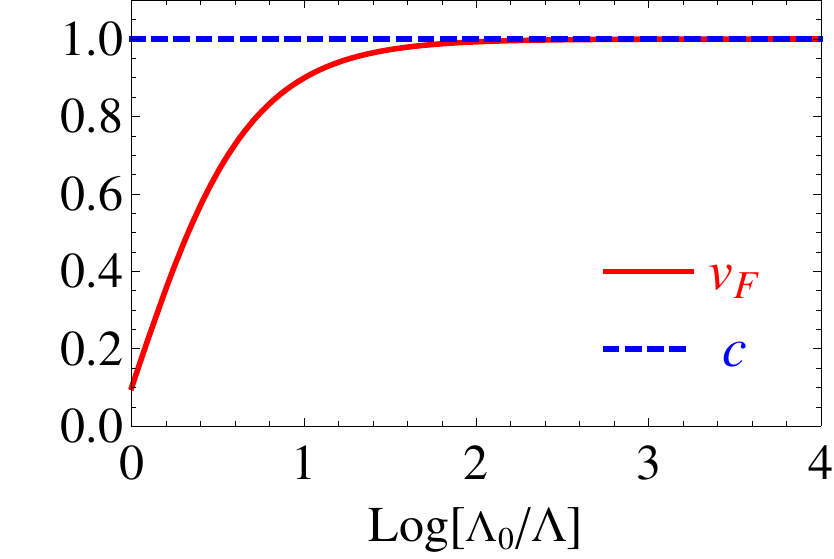}
\label{flow-charged-a}
}
\subfigure[]{
\includegraphics[scale=0.55]{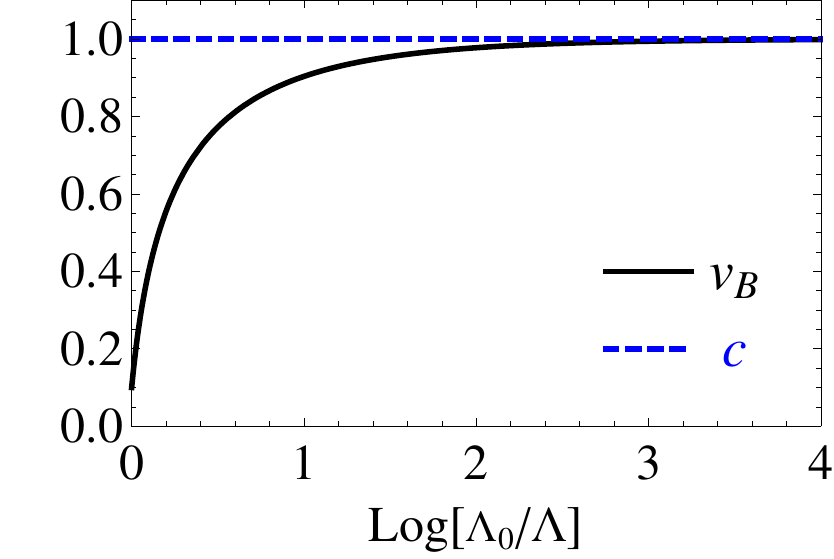}
\label{flow-charged-b}
}
\subfigure[]{
\includegraphics[scale=0.55]{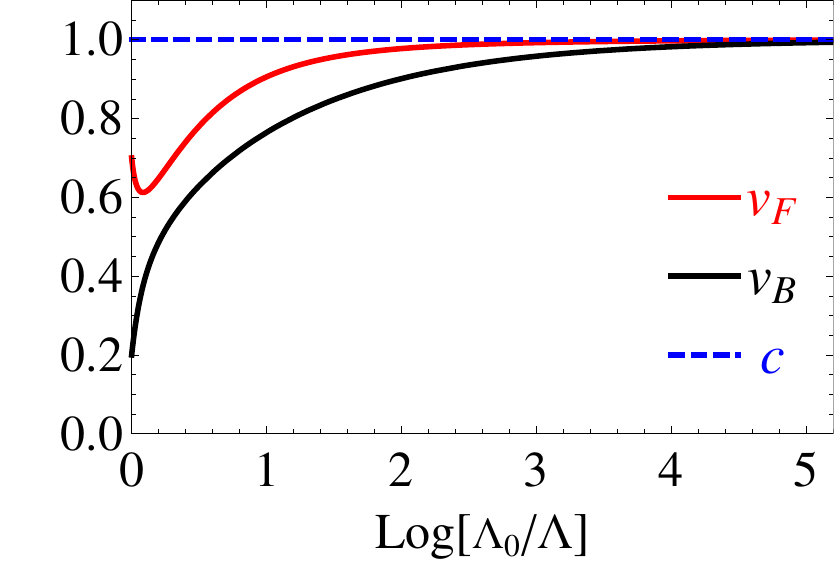}
\label{flow-charged-c}
}
\subfigure[]{
\includegraphics[scale=0.55]{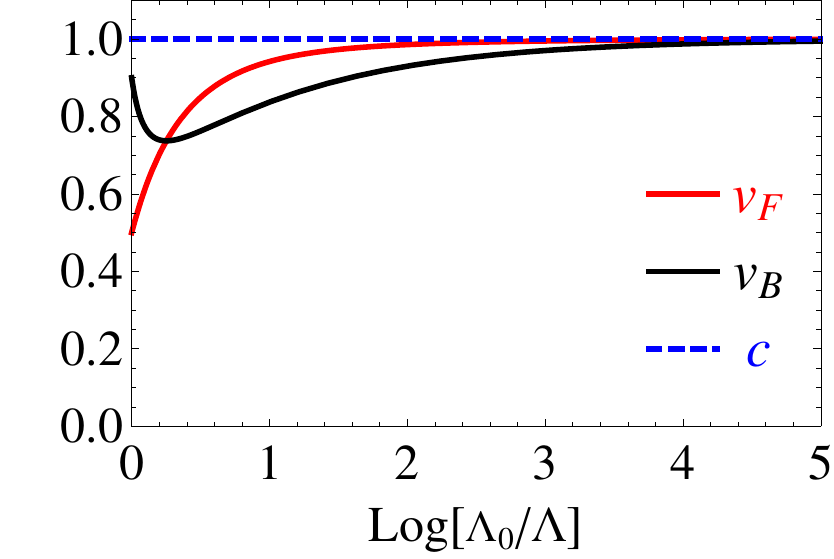}
\label{flow-charged-d}
}
\subfigure[]{
\includegraphics[scale=0.55]{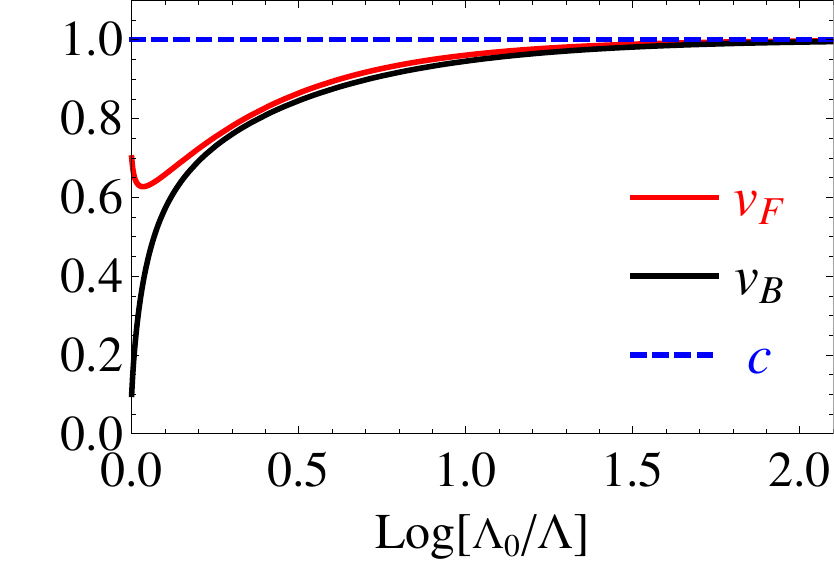}
\label{flow-charged-e}
}
\subfigure[]{
\includegraphics[scale=0.55]{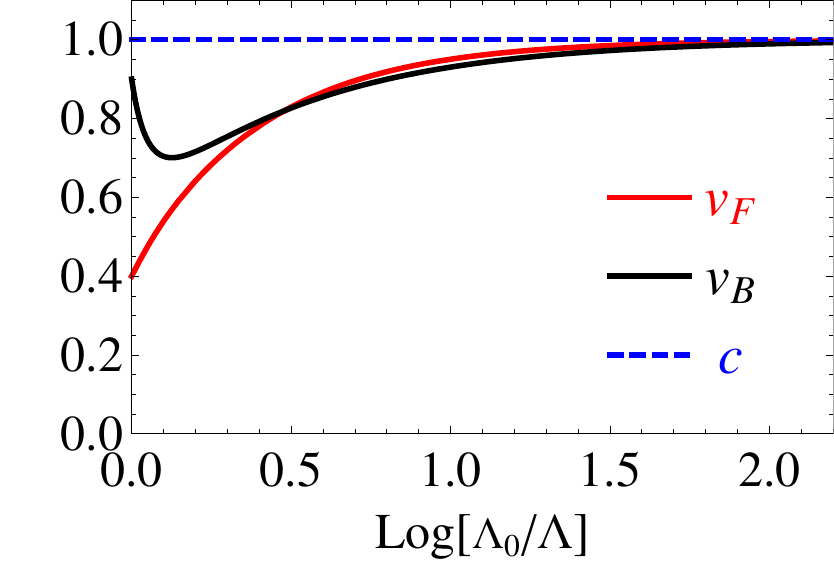}
\label{flow-charged-f}
}
\caption[]{ Renormalization group flows for (a) Fermi velocity (for $\alpha_F=1$), and (b) bosonic velocity ( for $\alpha_B=1$) for vanishing Yukawa coupling ($g=0$). Flows of $v_F$ (red) and $v_B$ (black) for $g=1, \alpha_F=1$, (c) $v^0_F>v^B_0$, and (d) $v^0_B<v^0_F$, when the bosonic field is assumed to be charge neutral. Flows of these two velocities for (e) $v^0_F>v^0_B$, and (f) $v^0_B>v^0_F$, when the bosonic field is also charged. Various parameters for this set of plots are $N_F=1$ [(a), (c)-(f)], and $g=1$, $\alpha_F=1$ [(c)-(f)]. Blue dotted lines represent the velocity of light ($c=1$), which in two dimensions does not renormalize. }\label{flow-charged}
\end{figure}

Next we systematically incorporate the coupling of fluctuating gauge fields with fermions and bosons. First, we consider a simple situation, when bosons and fermions are decoupled ($g=0$). Flow equations for $v_F$~\cite{vozmediano} and $v_B$ are then also decoupled, and with increasing RG time, both of them approach the velocity of light ($c$), however, at different rates, as shown in Fig.~\ref{flow-charged-a} and Fig.~\ref{flow-charged-b}, respectively. Notice that if we assume $v_F, v_B \ll c$ and neglect all the higher order terms $~\sim (v_a/c)^n$ for $a=F, B$ in the flow equations of $v_F$ and $v_B$, these two velocities continue to grow logarithmically with no upper bound, leading to a catastrophic outcome. However, as $v_a \to c$, one looses the liberty of working with only the instantaneous (density-density or Coulomb) interactions with the gauge field and neglect the current-current interaction, by virtue of which ultimately $v_a \to c$ and the causality in the system gets restored.

We now focus on the situation when massless fermions are coupled to fluctuating gauge fields, but the bosonic field is charge neutral. This scenario can be germane in the vicinity of a quantum phase transition out of Dirac semimetal into an insulating phase that spontaneously breaks an $O(2)$ rotational symmetry. One prototypical example of such insulating phase is the Kekule valence bond insulator in monolayer graphene. This situation can be analyzed by setting $e=0$ in Eq.~(\ref{boson-flow}). Numerically solving the flow equations we find that the terminal velocity for fermionic and bosonic excitations is the velocity of light $v_t=1(=c)$, irrespective of the ratio of the bare velocities, i.e., whether $v^0_F>v^0_B$ [see Figs.~\ref{flow-charged-c}] or $v^0_B>v_F^0$ [see Figs.~\ref{flow-charged-d}]. This outcome can be readily understood as follows. When fermions are coupled with gauge field, Fermi velocity ultimately approaches the velocity of light. Meanwhile, fermions give boost to bosonic velocity through Yukawa coupling ($g$), so that also $v_B \to v_F$. Thus, in the deep infrared regime both velocities meet each other and reach the velocity of light $v_F,v_B \to v_t=c$.

Finally, we delve into the most general scenario, when both fermion and boson are coupled with the gauge field, and the flow of Fermi and bosonic velocities is thus determined by Eqs.~(\ref{fermion-flow}) and (\ref{boson-flow}), respectively. Such field theory describes quantum-critical behavior near a continuous phase transition out of Dirac semimetal into the spin-singlet s-wave superconductor in graphene ($N_F=2$), surface states of a strong $Z_2$ topological insulator ($N_F=1/2$) and surface states of a topological crystalline insulator ($N_F=2$). The RG flow for $v_F$ and $v_B$ is shown in Fig.~\ref{flow-charged-e} for $v^0_F>v^0_B$, and in Fig.~\ref{flow-charged-f} for $v^0_B>v^0_F$. Irrespective of their initial values, both $v_F$ and $v_B$ flow toward the velocity of light as more and more degrees of freedom are integrated out, and the terminal velocity in the system is the velocity of light, $v_t=c=1$.

It is worth pointing out a peculiar feature in the flow of $v_F$ and $v_B$, when either only fermions or both fermion and boson are coupled with the gauge field. Let us consider the situation when $v^0_F>v^0_B$. In the short RG time, while $v_F$ decreases, $v_B$ continues to increase monotonically. However, both $v_F$ and $v_B$ ultimately meet the velocity of light after sufficiently long RG time, as shown in Fig.~\ref{flow-charged-c} (for neutral boson) and Fig.~\ref{flow-charged-e} (for charged boson). A similar behavior in bosonic velocity ($v_B$) is also present when $v^0_B>v^0_F$, as shown in Fig.~\ref{flow-charged-d} (for neutral boson) and Fig.~\ref{flow-charged-f} (for charged boson). Such distinct behavior in short and long time scale can be understood in the following way. For short time scale both $v_F$ and $v_B \ll c(=1)$, and the Yukawa coupling ($g$) between fermions and bosons dominates over their coupling with the gauge field. As a result, initially these two velocities try to meet each other, leading to their short time scale behavior. However, as one of the velocities grows (depending on weather $v^0_F>v^0_B$ or vice-versa) the coupling with the gauge field becomes gradually more important and finally both of them reach a universal terminal velocity $v_t=c=1$. Therefore, short length scale physics is governed by the Yukawa coupling, while coupling with the fluctuating gauge field dominates in the deep infrared regime. In contrast, in the absence of Yukawa coupling both $v_F$ and $v_B$ increase monotonically and meet the velocity of light, once sufficiently large number of degrees of freedom is integrated out from the system [see Fig.~\ref{flow-charged-a} and Fig.~\ref{flow-charged-b}, respectively].

\subsection{Renormalization group flow of velocities in three dimensions}\label{RG-3D}

We now proceed with the analysis of the flows of the velocities in three spatial dimensions. These flows are quite different than in two dimensions, since in $D=3$ the gauge field propagator is analytic, and therefore both permittivity and permeability of the medium receive perturbative corrections from the diagrams in  Fig.~\ref{self-energy-diag} $(f)-(h)$. In other words, both $U(1)$ charge ($e$) and velocity of light ($c$) flow in the RG sense, as given by Eqs.~(\ref{alpha-flow}) and (\ref{c-flow}), respectively.

\begin{figure}[htbp]
\centering
\subfigure[]{
\includegraphics[scale=0.5]{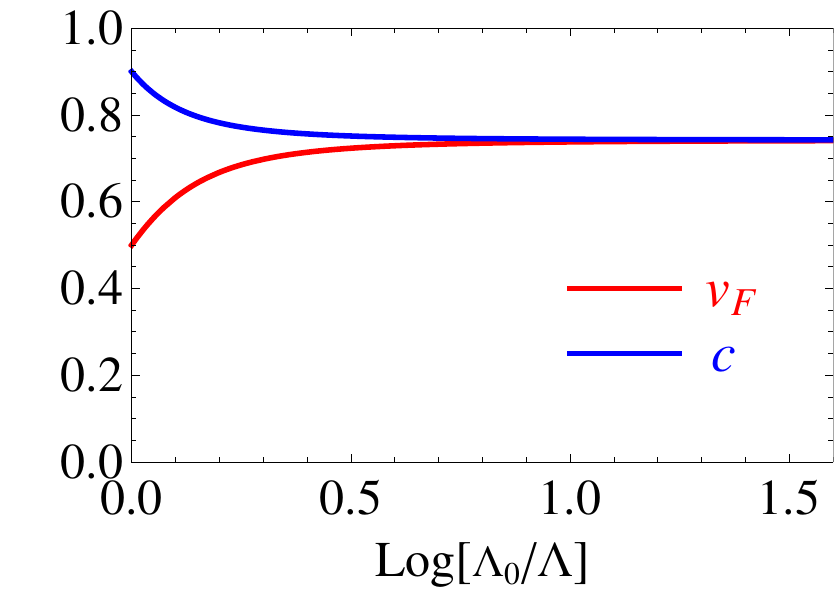}
\label{flow-charged-3D-a}
}
\subfigure[]{
\includegraphics[scale=0.5]{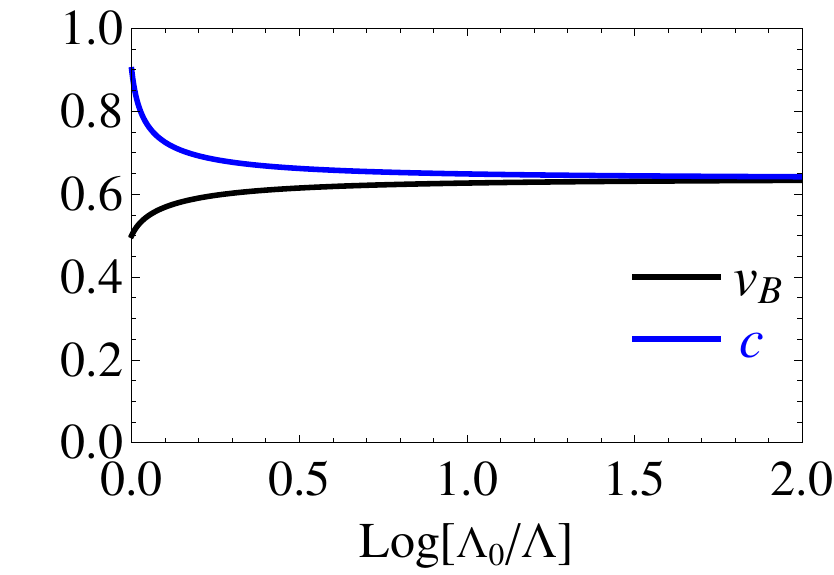}
\label{flow-charged-3D-b}
}
\subfigure[]{
\includegraphics[scale=0.5]{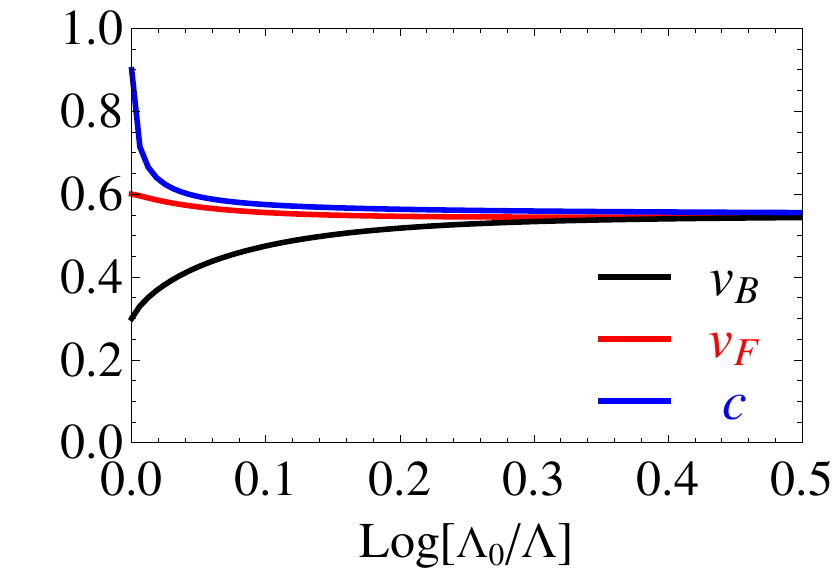}
\label{flow-charged-3D-c}
}
\subfigure[]{
\includegraphics[scale=0.5]{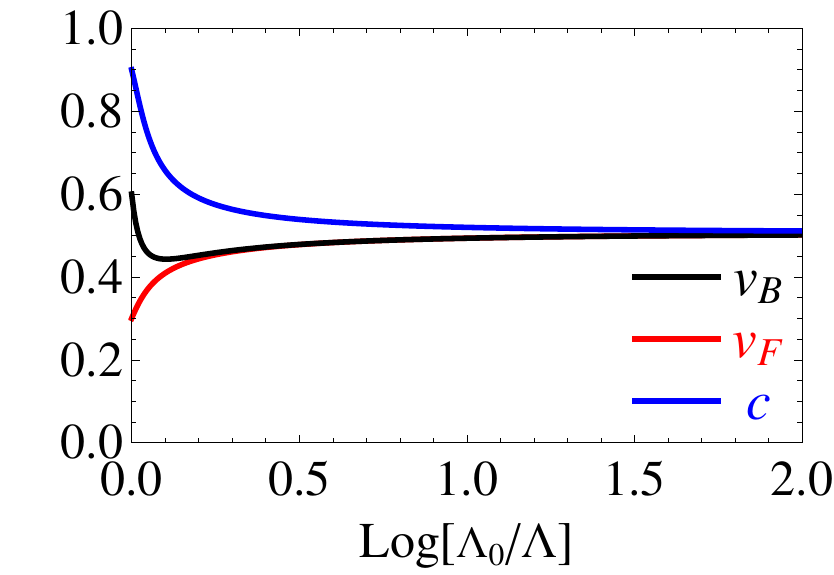}
\label{flow-charged-3D-d}
}
\subfigure[]{
\includegraphics[scale=0.5]{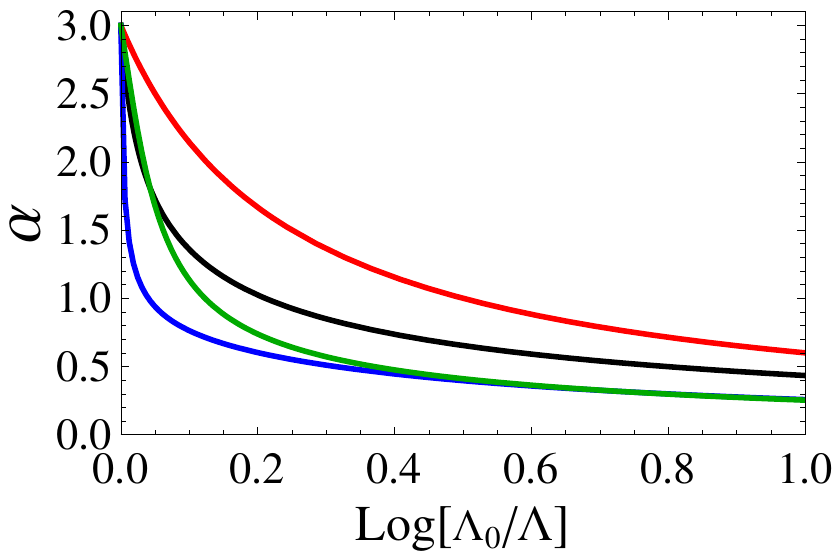}
\label{flow-charged-3D-e}
}
\subfigure[]{
\includegraphics[scale=0.5]{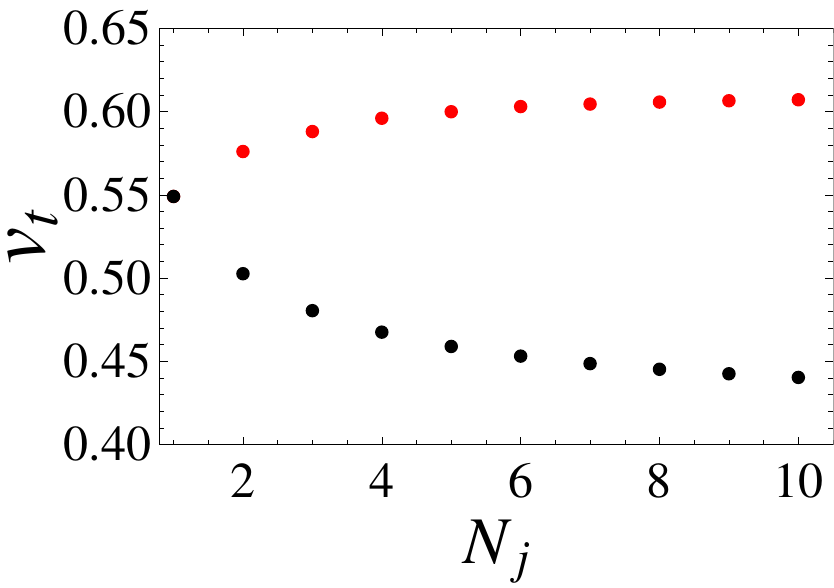}
\label{flow-charged-3D-f}
}
\caption[] {(a) Flow of $v_F$ (red) and $c$ (blue) for $v^0_F<c^0$, $N_F=1$ and $\alpha_F^0=3$ in a pure fermionic theory. (b) Flow of $v_B$ (black) and $c$ (blue) for $v^0_B<c^0$, $N_B=1$, and $\alpha_B^0=e^2/(\varepsilon c^2 v_B)=3$ in a pure charged bosonic theory. Flow of three velocities for (c) $v^0_B<v^0_F<c^0$, (d) $v^0_F<v^0_B<c^0$. In these two plots, we set $g=1$, $\alpha^0_F=3$, and $N_F=N_B=1$. (e) Respective for Fig.~(a)-(d), the flow of fine structure constant is shown in red ($\alpha_F$), black ($\alpha_B$), blue ($\alpha_F$) and green ($\alpha_F$). (f) Dependence of the terminal velocity ($v_t$) on fermionic flavor number $N_F$ for fixed $N_B=1$ (red dots), and bosonic flavor number $N_B$ for fixed $N_F=1$, when $v^0_B<v^0_F<c^0$, $\alpha^0_F=3$, and $g=1$.} \label{flow-charged-3D}
\end{figure}

If we neglect the presence of bosonic field, flow of $v_F$ and $c$ is shown in Fig.~\ref{flow-charged-3D-a}. Corresponding flow of the fine structure constant is shown in Fig.~\ref{flow-charged-3D-e} in red. Therefore, as system approaches deep infrared regime, $v_F$ ($c$) logarithmically increases (decreases), and ultimately two velocities acquire a common, but non-universal value $v^0_F<v_t<c^0$, while $\alpha_F$ continues to decrease monotonically~\cite{isobe}. If we neglect the terms $\sim (v_F/c)^n$ (situation with only the instantaneous piece of the gauge-fermion interaction), the Fermi velocity continues to increase monotonically~\cite{goswami-chakravarty, gonzalez, dassarma}. However, as $v_F \to c$, the retarted components of the interactions become comparable with the instantaneous one, and the growth of $v_F$ stops. Ultimately the system is described by an unique terminal velocity ($v_t$), where $v^0_F<v_t<c^0$. Similar outcome is found if we consider a system of charged bosons in the absence of fermions. The bosonic velocity and velocity of light meet as the system approaches infrared, as shown in Fig.~\ref{flow-charged-3D-b}. Ultimately, the photon and boson are described by a common non-universal terminal velocity $v_t$, where $v^0_B<v_t<c^0$. The fine structure constant in such a system, defined as $\alpha_B=e^2/(\varepsilon c^2 v_B)$, decreases monotonically [see black curve in Fig.~\ref{flow-charged-3D-e}].

Finally, we consider the boson-fermion coupled system in three dimensions, when these two degrees of freedom are also coupled with fluctuating gauge field. The flow of $v_F$ (red), $v_B$ (black) and $c$ (blue) are shown in Fig.~\ref{flow-charged-3D-c} and Fig.~\ref{flow-charged-3D-d}, respectively, for $v^0_B<v^0_F<c^0$ and $v^0_F<v^0_B<c^0$. Therefore, irrespective of the initial velocity of fermionic and bosonic degrees of freedom, ultimately all three velocities acquire a common, but non-universal value. In both situations the fermionic  fine structure constant, $\alpha_F$, decreases monotonically as more and more degrees of freedom are integrated out. The flows of $\alpha_F$ in these two situations are shown in blue and green in Fig.~\ref{flow-charged-3D-e}, respectively. For the sake of simplicity, we here do not discuss the situation when bosonic field is charge-neutral, but Dirac fermions are coupled to the gauge field. The outcome, however, remains the same that all three velocities finally acquires a common value~\cite{anber}.

The terminal velocity in boson-fermion-photon coupled theory displays an interesting dependence on the fermionic ($N_F$) and the bosonic ($N_B$) flavor numbers. For a fixed $N_B$, as the number of fermionic degrees of freedom increases, the terminal velocity increases monotonically, but asymptotically saturates to a value such that $v^0_F<v_t<c^0$ in the limit $N_F \to \infty$, at least when $v^0_F>v^0_B$. Such behavior arises from the fact that with increasing fermion flavor number, $v_B$ receives a larger boost (when $v^0_B<v^0_F$), and the terminal velocity increases monotonically. On the other hand, for fixed $N_F$, when bosonic flavor number increases, the terminal velocity decreases monotonically and asymptotically saturates to a value such that $v^0_B<v_t<v^0_F,c^0$ as $N_B \to \infty$, for the bare velocities $v^0_F>v^0_B$. The difference in behaviors originates in the appearance of $N_B$ in the flow equation of $c$ [see Eq.~(\ref{c-flow})]. With increasing bosonic flavor number the velocity of light decreases faster in the medium, and consequently the terminal velocity decreases monotonically. In both cases the velocity of the degree of freedom with the larger number of species sets the lower bound for the terminal velocity. Such dependence of the terminal velocity on flavor number ($N_F$ and $N_B$) is shown in Fig.~\ref{flow-charged-3D-f} for $v^0_F>v^0_B$.

Therefore, irrespective of dimensionality of the system and the bare values of the velocities, in the ultimate infrared regime all degrees of freedom acquire a common velocity, ensuring the Lorentz symmetric fixed point as the ultimate stable fixed point in the theory.

\section{Conclusions}\label{Conclusion}

In conclusion, we have here studied the emergence of the Lorentz symmetry in the field theory describing the coupling of the massless fermions to the bosons through the Yukawa coupling. These two degrees of freedom, in addition, are coupled to a fluctuating $U(1)$ gauge field in two and three spatial dimensions. This theory in its different variations describes quantum-critical behavior of interacting Dirac fermions in various condensed-matter systems. Our analysis is performed within one-loop $\epsilon-$expansion about three spatial dimensions, which is the upper critical dimension in the problem. In two spatial dimensions, both the charge and the velocity of light are protected against perturbative corrections, due to the non-analyticity of the gauge field propagator and the gauge symmetry, and we obtain that both fermionic and bosonic velocities ultimately flow to the velocity of light, albeit at different rates. Interestingly, even when fermions and bosons are decoupled from the gauge field, they also reach a common terminal velocity, and Lorentz symmetry is restored in the infrared in this case as well. The terminal velocity, however, depends on both the bare velocities and the number of the flavors, and in that sense is non-universal. Therefore, only when the gauge coupling is turned on, the terminal velocity in 2+1 dimensions is completely universal and set by the velocity of the $U(1)$ gauge field, i.e., the velocity of light ($c$). Finally, we would like to emphasize that the emergence of the Lorentz symmetry is completely independent of the value of the Yukawa coupling, implying that the strongly coupled  quantum-critical points governing various phase transitions of Dirac fermions  are all Lorentz invariant. This was previously hinted by performing a stability analysis of the assumed Lorentz symmetry at the quantum-critical points~\cite{HJR, HJV}, but the analysis performed here shows that the Lorentz symmetry emerges at such quantum-critical points for arbitrary values of the bare velocities.

In three spatial dimensions both the charge and the velocity of light flow, and this leads to a richer behavior of the velocities  as compared to the situation in two dimensional systems. First of all, even in the theory with charged fermions or bosons only, the velocity flows to a common terminal velocity, which is however, non-universal, as it depends on both the number of the flavors and the initial conditions. In the theory with both fermions and bosons charged and also coupled through the Yukawa coupling, all three velocities flow to a common terminal velocity, which is also non-universal, and with its lower bound set by the initial velocity of the degrees of freedom (fermions or bosons) with less number of flavors. Therefore, emergence of the Lorentz symmetry in the infrared seems to be a universal feature of this theory in three dimensions, at least within our one-loop RG analysis~\cite{HJR}.

Although we have illustrated the universal phenomena regarding the flow of the velocities in a system of interacting massless Dirac fermions residing in the close proximity to an ordering that breaks continuous $O(2)$ symmetry, the obtained results can be generalized to various order parameters breaking other symmetries. For example, one can focus near the quantum phase transition of quasi-relativistic fermions towards the formation of an ordered state breaking either a discrete $Z_2$ or continuous $O(3)$ symmetry. In graphene, such theories respectively describe the transition to a charge-density-wave and N\'{e}el antiferromagnetic orderings~\cite{HJV}. When Dirac fermions are additionally coupled with a $U(1)$ gauge field, while the bosonic field remains charge neutral, our results imply that at either of these two quantum critical points, the ultimate terminal velocity is the velocity of light. Our analysis is also germane near \emph{multi-critical} points in a system of strongly interacting Dirac fermions~\cite{broy-multi, roy-juricic, classen}. At a multi-critical point two distinct order parameter condense simultaneously and a \emph{super-order-parameter} with a larger symmetry is realized. Two bosonic fields then also have different bare velocities. Although we have not carried out the calculations specific to this situation, the present analysis strongly suggests that at the multi-critical point two bosonic fields should acquire a common velocity, which due to the Yukawa coupling to Dirac fermions ultimately flows to the velocity of light.

As the penultimate remark, we should point out that as far as the anisotropy of velocity of different degrees of freedom is concerned, the outcomes in two and three dimensions are qualitatively the same. However, the quantum-critical behavior in two and three spatial dimensions is completely distinct. While the transition out of Dirac semimetal into an ordered phase in 2+1 dimensions is non-mean-field in nature and generically occurs at strong coupling, its (3+1)-dimensional counterpart is only Gaussian (mean-field). This stems from the fact that $D=3$ is the upper-critical dimension for the instabilities of massless Dirac fermions. Hence, correlation length exponent $\nu$ acquires a mean-field value ($\nu=1/2$) and the hyperscaling hypothesis is violated~\cite{zinn-justin}.

Finally, we comment on the crossover behavior of some transport and thermodynamic quantities when the bare velocities of matter fields (fermionic, bosonic) are different (and typically much smaller) than the velocity of light. For example, in two and three spatial dimensions the specific heat of massless Dirac fermions scales as $C_v \sim T^2/v^2_F$ and $T^3/v^3_F$, respectively. When relativistic fermions are coupled only to photon, $v_F$ increases logarithmically toward the velocity of light $c$. Consequently, $C_v/ T^n $ suffers a monotonic decrease as the temperature in gradually decreased. The zero-temperature optical or frequency ($\Omega$) dependent conductivity in three-dimensional quasi-relativistic system scales as $\sigma(\Omega) \sim \Omega/v$, which thus also decreases as the frequency in lowered.
In contrast, when fermions are coupled to bosonic (either charged or neutral) degrees of freedom, and in particular when their bare velocities are such that $v^{0}_F>v^{0}_B$, (Figs.~\ref{flow-charged-c} and ~\ref{flow-charged-e} for two dimensional system, or in Fig.~\ref{flow-charged-3D-c} for three dimensional systems), $v_F$ displays a \emph{nonmonotonic} behavior as system approaches the infrared sector, but ultimately reaches the velocity of light ($c$).
As a consequence, both thermodynamic (such as specific heat) and transport observables (such as conductivity in $D=3$) display nonmonotonic behavior. For example, specific heat in $D=2$ and $3$ increases initially, but turns around at intermediate (non-universal) temperature to ultimately saturate to a smaller value. Therefore, in a sufficiently clean Dirac or Weyl materials various measurable quantities can display such crossover phenomena.

\acknowledgments

B. R. is in debt to Pallab Goswami for valuable discussions. B. R. was suported by NFS-JQI-PFC and LPS-CMTC. V. J. acknowledges financial support from Netherlands Organization for Scientific Research (NWO). I. F. H. was supported by NSERC of Canada. We thank Max Planck Institute for the Physics of Complex Systems, Dresden for hospitality during the ``Advance Study Group" (2013) program, where this work was initiated. B. R. is thankful to Aspen Center of Physics for hospitality during the Summer Program (2015), where this work was finalized.

\appendix

\section{Fermionic self energy}\label{append-fermion-SE}

In this Appendix we show the evaluation of self-energy corrections for massless Dirac fermions due to its coupling with the fluctuating gauge field and the bosonic order parameter field through the Yukawa coupling. To the one-loop order the self energy corrections of Dirac fermions arise from the diagrams Figs.~\ref{self-energy-diag} $(a)$ and $(b)$. Contribution from Fig.~\ref{self-energy-diag} $(a)$ reads as
\begin{eqnarray}
(1a) &=& -\frac{e^2}{\varepsilon} \int \frac{d^{D}p}{(2 \pi)^D} \int^{\infty}_{-\infty} \frac{d \omega}{2 \pi} \: \gamma_\mu \gamma_5 \frac{i \left[ (\omega+\nu) \gamma_0 + v_F (p_j+k_j) \gamma_j  \right]}{(\omega+\nu)^2 +v^2_F (p+k)^2} \: \gamma_\nu \gamma_5 \: \frac{\delta_{\mu \nu}}{\omega^2+c^2 p^2} \nonumber \\
&=& -\frac{i e^2}{4 \varepsilon} \int^1_0 dx \int \frac{d^{D}p}{(2 \pi)^D} \: \gamma_\mu \: \left[ \frac{(1-x) \nu \gamma_0 + v_F(p_j+k_j) \gamma_j}{ [c^2 (1-x)+x v^2_F]^{3/2}} \right]\: \gamma_\nu \nonumber \\
&\times& \frac{1}{\left[ \left(p+ \frac{x v^2_F}{c^2(1-x)+x v^2_F} k\right)^2 + k^2 \frac{x(1-x)c^2 v^2_F}{[c^2(1-x)+xv^2_F]^2}+ \nu^2 \frac{x(1-x)}{c^2(1-x)+xv^2_F}\right]^{3/2}},
\end{eqnarray}
after performing the integration over the Matsubara frequency ($\omega$) and introducing the Feynman parameter ($x$), where $\gamma_\mu \equiv \left( \gamma_0, \frac{v_F}{c} \gamma_j \right)$. Performing a shift in variable $p+ \frac{x v^2_F}{c^2(1-x)+x v^2_F} k \to p$, and the integral over the momentum ($p$) in dimensions $D=3-\epsilon$, we obtain the divergent piece of the fermionic self-energy
\begin{eqnarray}\label{Ferm-SE-charge}
(1a)=-\frac{e^2}{4 \pi^2 \varepsilon} \; \left[ \frac{i \nu \gamma_0}{c(c+v_F)^2} \left( 1-3 \frac{v^2_F}{c^2} \right) -\frac{i v_F \gamma_j k_j \: (2 c+v_F)}{3 c v_F (c+v_F)^2} \left( 1+ \frac{v^2_F}{c^2} \right) \right] \; \frac{1}{\epsilon} + {\mathcal O} (1),
\end{eqnarray}
after performing the integral over $x$.

Self-energy correction from Fig.~\ref{self-energy-diag} $(b)$ reads
\begin{eqnarray}
&&(1b)= g^2 \int \frac{d^{D}p}{(2 \pi)^D} \int^{\infty}_{-\infty} \frac{d \omega}{2 \pi} \left[  \frac{P_+ \left[ i \left( \omega \gamma_0 +v_F \gamma_j p_j \right) \right]P_-}{[\omega^2 + v^2_F p^2][(\nu-\omega)^2 +v^2_B (k-p)^2 +m^2]}  + \left( P_+ \leftrightarrow P_- \right)\right] \nonumber \\
&=& \frac{g^2}{4} \int^1_0 dx \int \frac{d^{D}p}{(2 \pi)^D} \frac{i \gamma_0 x \nu + i \gamma_j v_F \left( p_j + \frac{x v^2_B}{v^2_F(1-x) +x v^2_B}k_j \right) [v^2_F(1-x) +x v^2_B]^{-3/2}}{\left[ p^2+ \frac{x(1-x) v^2_B v^2_F }{(v^2_F(1-x) +x v^2_B)^2} k^2 + \frac{x(1-x)\nu^2 +(1-x)m^2}{v^2_F(1-x) +x v^2_B} \right]^{3/2}},
\end{eqnarray}
where $P_\pm=\frac{1}{2} \left(1 \pm \gamma_5 \right)$ after completing the integration over the frequency, and shifting the momentum $ p$ $-\frac{x v^2_B}{v^2_F(1-x) +x v^2_B} k$ $\to p$. Performing the integration over momentum ($p$) in $D=3-\epsilon$ and over $x$, we obtain
\begin{eqnarray} \label{Ferm-SE-Yukawa}
(1f)=\frac{g^2}{4 \pi^2 \epsilon} \: \frac{1}{v_B (v_F+v_B)^2}  \left[ i \nu \gamma_0 + i \gamma_j v_F k_j \:\: \left( \frac{1+2 \; \frac{v_B}{v_F}}{3} \right) \right] + {\mathcal O} (1).
\end{eqnarray}
Combining the expressions from Eqs.~(\ref{Ferm-SE-charge}) and (\ref{Ferm-SE-Yukawa}), we finally obtain the expressions for fermionic field renormalization ($Z_\Psi$) in Eq.~(\ref{field-fermion}) and renormalization of Fermi velocity ($Z_{v_F}$) in Eq.~(\ref{velocity-fermion}).

\section{Bosonic self-energy}\label{append-boson-SE}

This Appendix is devoted to the evaluation of self-energy corrections of bosons carrying $U(1)$ charge $2e$  due to their coupling to the fluctuating gauge field and  to massless Dirac fermions through the Yukawa coupling. Correction to the bosonic self-energy due to Yukawa coupling arise from the diagram shown in Fig.~\ref{self-energy-diag} $(c)$ and its contribution is given by
\begin{eqnarray}
&(1c)&= 2 g^2 N_F \int \frac{d^{D}q}{(2 \pi)^D} \int^{\infty}_{-\infty} \frac{d \omega}{2 \pi} \:
\frac{\omega (\omega+\nu) + v^2_F \vec{q} \cdot (\vec{p}+\vec{q})}{ (\omega^2 + v^2_F q^2) \left[ (\omega+\nu)^2 + v^2_F (p+q)^2 \right] } = \frac{g^2 N_F}{2} \int^1_0 dx  \nonumber \\
&\times& \int \frac{d^{D}q}{(2 \pi)^D} \bigg[ \frac{v^2_F q^2 + x(1-x) (\nu^2+v^2_F p^2) -x(1-x) \nu^2 +v^2_F q^2 -x(1-x) v^2_F p^2}{\left[ v^2_F q^2 + x(1-x) (\nu^2+v^2_F p^2) \right]^{3/2} },
\end{eqnarray}
after completing the frequency integral. Performing the integration over the momentum we arrive at
\begin{equation}\label{boson-yukawa}
(1c)=-\frac{g^2 N_F}{8 \pi^2 v^3_F} \: \left( \nu^2 + v^2_F p^2 \right) \; \frac{1}{\epsilon} + {\mathcal O} (1).
\end{equation}
On the other hand, correction to bosonic self-energy due to the fluctuating gauge field [see Fig.~\ref{self-energy-diag} $(d)$] is given by
\begin{eqnarray}
(1d)&=& \frac{4 e^2}{\varepsilon} \int \frac{d^{D}p}{(2 \pi)^D} \int^{\infty}_{-\infty} \frac{d \omega}{2 \pi} \frac{1}{\omega^2 + c^2 p^2}
\: \frac{(2 \nu+\omega)^2 +v^4_B \; c^{-2} (2 k +p)^2}{ (\omega+\nu)^2 +v^2_B (k+p)^2 +m^2} \nonumber \\
&=& \frac{e^2}{\varepsilon} \int^1_0 dx  \int \frac{d^{D}q}{(2 \pi)^D} \bigg[ \frac{[c^2(1-x)+xv^2_B]^{-1/2}}{\left[ \left( p+\frac{x v^2_B}{c^2(1-x)+x v^2_B} k\right)^2 + \frac{x(1-x) v^2_B c^2}{[c^2(1-x)+x v^2_B]^2} k^2 + \frac{x(1-x)\nu^2 +x m^2}{c^2(1-x)+x v^2_B} \right]^{1/2} } \nonumber \\
&+& \frac{\left[ (c^2(1-x)+xv^2_B)\right]^{-3/2} \left[(2-x)^2 \nu^2 +v^4_B \; c^{-2} (2k+p)^2 \right]}{\left[ \left( p+\frac{x v^2_B}{c^2(1-x)+x v^2_B} k \right)^2 + \frac{x(1-x) c^2  v^2_B}{(c^2(1-x) +x v^2_B)^2} k^2 + \frac{x(1-x)\nu^2 +x m^2}{c^2(1-x)+x v^2_B} \right]^{3/2} }
\bigg],
\end{eqnarray}
after completing the frequency integral. Taking $p+\frac{x v^2_B}{1-x+x v^2_B} k \to p$, and performing the momentum integral in $D=3-\epsilon$ dimension we obtain
\begin{eqnarray}
(1d)=\frac{e^2}{\pi^2 \varepsilon} \bigg[ \frac{c^2+2 c v_B -v^2_B}{c^3 v_B (c+v_B)} \; \nu^2 + \frac{38 c^3+31 c^2 v_B -6 c v^2_B-3 v^3_B}{15 c^3 v_B \left(c+v_B \right)^2} \; v^2_B k^2 \bigg] \; \frac{1}{\epsilon} +{\mathcal O}(1). \label{boson-charge}
\end{eqnarray}

Finally, the contribution from Fig.~\ref{self-energy-diag} $(e)$ reads
\begin{eqnarray}
(1e) = \frac{\lambda}{2} \left( 2 N_B +2 \right) \int \frac{d^{D}q}{(2 \pi)^D} \int^{\infty}_{-\infty} \frac{d\omega}{2 \pi} \: \frac{1}{\omega^2+v^2_B q^2 +m^2}
=  \frac{\lambda (N_B+1)}{v^3_B} \; m^2 \; \frac{1}{\epsilon} + {\mathcal O} (1),
\end{eqnarray}
after taking $\lambda/(8 \pi^2) \to \lambda$, and gives renormalization to bosonic mass only. Upon combining the contributions from Eqs.~(\ref{boson-yukawa}) and (\ref{boson-charge}), we find the normalization of bosonic field ($Z_\Phi$) [ see Eq.~(\ref{field-boson})] and bosonic velocity ($Z_{v_B}$) [see Eq.~(\ref{velocity-boson})].

\section{Self-energy correction of gauge field}\label{gauge-SE-appendix}

This Appendix is devoted to the computation of the renormalization of the gauge field propagator in three spatial dimensions. We first focus on such corrections due to the gauge-fermion coupling. The contribution from the diagram $(f)$ in Fig.~\ref{self-energy-diag} reads as
\begin{equation}\label{gauge-fermion}
(1f)=e^2 \mbox{Tr} \int^{\infty}_{-\infty} \frac{d\omega}{2\pi}\int \frac{d^D q}{(2\pi)^D}
\frac{\gamma_\mu ({\slashed k} + \slashed{q}) \gamma_\nu \slashed{q}}{(k+q)^2 q^2}
=\frac{e^2}{v^D_F}  \mbox{Tr}\int \frac{d^{D+1} q}{(2\pi)^{D+1}} \frac{\gamma_\mu ( \slashed{k} + \slashed{q}) \gamma_\nu \slashed{q}}{(k+q)^2 q^2}.
\end{equation}
Since
\begin{equation}
\slashed{k}+\slashed{q}=\gamma_0 (\omega+\nu)+ v_F \gamma_j (k+q)_j, \:
\slashed{q}=\gamma_0 \omega + v_F \gamma_j q_j
\end{equation}
we perform a change in variable $v_F q_j \to q_j$. After this transformation the frequency and momentum integrals in Eq.~(\ref{gauge-fermion}) can be promoted to $d=D+1$-dimensions, and we can extract the divergent piece in the above expression by performing the integration using $d=4-\epsilon$ scheme. The result is
\begin{equation}
(1f)=\frac{4}{3} e^2 N_F \: \left( k^2 \delta_{\mu \nu}- k_\mu k_\nu \right) \: \left( \frac{v_F}{c}\right)^{2-\delta_{\mu,0}-\delta_{\nu,0}} \; \frac{k^{\epsilon}}{v^3_F \epsilon} +{\mathcal O} (1),
\end{equation}
after taking $e^2/(8 \pi^2) \to e^2$, since $\gamma_\mu=\left(\gamma_0, \frac{v_F}{c} \gamma_j \right)$, where $k_\mu=(k_0,v_F k_j)$. To extract the renormalization of permittivity ($\varepsilon$) and permeability ($\mu$) of the medium we need to rescale the time coordinate as $c \tau \to x_0$ in the action. Under this rescaling $k_\mu=(k_0,v_F k_j)/c$. Finally promoting the theory from imaginary to real time (Wick rotation), we obtain
\begin{equation}
(1f)= \frac{1}{2} \: A_\mu \left[ \frac{4}{3} e^2 N_F \: \left( k^2 \; {\mathcal G}^{\mu \nu}- k^\mu k^\nu \right) \: \left( \frac{v_F}{c}\right)^{2-\delta_{\mu,0}-\delta_{\nu,0}} \; \frac{k^{\epsilon}}{v^3_F} \right] A_\nu \; \frac{1}{\epsilon} +{\mathcal O}(1),
\end{equation}
where ${\mathcal G}^{\mu \nu}=\mbox{Diag.} \left(1,-1,-1,-1 \right)$.

Next we evaluate the renormalization of the gauge field propagator through its coupling with the bosonic order parameter field. The relevant diagrams to one loop order are shown in Fig.~\ref{self-energy-diag} $(g)$ and $(h)$. Contribution from the diagram $(g)$ reads as
\begin{eqnarray}
(1g)= N_B (2e)^2 \int^{\infty}_{-\infty} \frac{d\omega}{2 \pi} \int \frac{d^D q}{(2 \pi)^D} \frac{(k+2 q)_\mu (k+2 q)_\nu}{\left(q^2+m^2 \right) \left[ (q+k)^2 + m^2\right]},
\end{eqnarray}
where $(k+2 q)_\mu= \left( \nu+2 \omega, v^2_B (k+2q)_j/c \right)$. After taking $v^2_B q_j/c \to q_j$, we once again can promote the above integral to $d=D+1$ dimensional space-time (imaginary) integration. Performing the integral using $d=4-\epsilon$ scheme we obtain
\begin{eqnarray}
(1g) &=& \frac{4 N_B e^2}{\left( v^2_B/c\right)^3} \int \frac{d^{d} q}{(2 \pi)^{d}} \frac{(k+2 q)_\mu (k+2 q)_\nu}{\left(q^2+m^2 \right) \left[ (q+k)^2 + m^2\right]} \nonumber \\
&=& \left[ \frac{4 N_B e^2}{3 \left( v^2_B/c\right)^3} \left[ k^2 \delta_{\mu \nu}-k_\mu k_\nu \right] \; \left( \frac{v_B}{c} \right)^{2-\delta_{\mu,0}-\delta_{\nu,0}} + \frac{8 N_B e^2}{\left( v^2_B/c\right)^3 } m^2 \delta_{\mu \nu} \right] \; \frac{1}{\epsilon}, \label{boson-bubble}
\end{eqnarray}
after taking $e^2/(8 \pi^2) \to e^2$, where $k_\mu=\left( k_0, v^2_B k_j/c\right)$. The second term in the last equation cancels with the contribution from diagram $(h)$ in Fig.~\ref{self-energy-diag} and the gauge field remains transverse. So, from now on we focus on the first term of Eq.~(\ref{boson-bubble}). After rescaling the imaginary time according to $c \tau \to x_0$, we obtain $k_\mu=\left( k_0/c, v^2_B k_j/c^2\right)$. Finally, changing the imaginary time to real one (Wick rotation), we obtain
\begin{equation}
(1g)+(1h)= \frac{1}{2} \: A_\mu \left[ \frac{4}{3} e^2 N_B \: \left( k^2 \; {\mathcal G}^{\mu \nu}- k^\mu k^\nu \right) \: \left( \frac{v_B}{c}\right)^{2-\delta_{\mu,0}-\delta_{\nu,0}} \; \frac{k^{\epsilon}}{\left( v^2_B/c \right)^3} \right] A_\nu \frac{1}{\epsilon} +{\mathcal O}(1).
\end{equation}

Hence, the total correction to gauge field propagator reads as
\begin{eqnarray}
\Pi_2 &=& \frac{1}{\mu c^2}\bigg[ A^2_1 \left( k^2_2+k^2_3\right) + A^2_2 \left( k^2_3+k^2_1\right)+A^2_3 \left( k^2_1+k^2_2\right) -2 A_j A_l k_j k_l \bigg] \left[\frac{v_F}{c^2} \Gamma^{\mu}_F  +\frac{c}{v^2_B} \Gamma^{\mu}_B \right] \nonumber \\
&+& \varepsilon \bigg[-A^2 \frac{k^2_0}{c^2} -A^2_0 k^2 -2 A_0 A_j \frac{k_0}{c} k_j \bigg] \left[ \frac{1}{c^2 v_F}\Gamma^{\varepsilon}_F+\frac{c}{v^4_B} \Gamma^{\varepsilon}_B \right],
\end{eqnarray}
where
\begin{equation}
\Gamma^{\varepsilon}_F =\frac{4}{3 \varepsilon} N_F e^2 \frac{k^{-\epsilon}}{\epsilon},\:
\Gamma^{\mu}_F =\frac{4 \mu}{3} N_F e^2 \frac{k^{-\epsilon}}{\epsilon},\:
\Gamma^{\varepsilon}_B =\frac{4}{3 \varepsilon} N_B e^2 \frac{k^{-\epsilon}}{\epsilon},\:
\Gamma^{\mu}_F =\frac{4 \mu}{3} N_B e^2 \frac{k^{-\epsilon}}{\epsilon},
\end{equation}
and summation over repeated indices is assumed. In real time the Maxwell Lagrangian reads as
\begin{eqnarray}\label{em-realtime}
L_{EM} &=&\varepsilon E^2-\frac{1}{\mu} B^2 = \varepsilon \bigg[-A^2 \frac{k^2_0}{c^2} -A^2_0 k^2 -2 A_0 A_j \frac{k_0}{c} k_j \bigg] \nonumber \\
&-& \frac{1}{\mu c^2}\bigg[ A^2_1 \left( k^2_2+k^2_3\right) + A^2_2 \left( k^2_3+k^2_1\right)+A^2_3 \left( k^2_1+k^2_2\right) -2 A_j A_l k_j k_l \bigg],
\end{eqnarray}
since
\begin{equation}
\vec{E}=-\frac{1}{c} \frac{\partial \vec{A}}{\partial t}- \vec{\nabla} A_0, \quad
\vec{B}=\frac{1}{c} \vec{\nabla} \times \vec{A}.
\end{equation}
From the above equations we then arrive at the renormalization conditions for $\varepsilon$ and $\mu$
\begin{eqnarray}
Z_{\varepsilon^{-1}}=1-\frac{4e^2}{3 \varepsilon} \left( \frac{N_F}{c^2 v_F} + N_B \frac{c}{v^4_B} \right) \frac{1}{\epsilon}, \quad
Z_{\mu^{-1}}=1 + \frac{4e^2}{3 \varepsilon} \left( N_F \frac{v_F}{c^2} + N_B \frac{c}{v^2_B} \right) \frac{\varepsilon \mu}{\epsilon}.
\end{eqnarray}
Since $c^2=1/(\varepsilon \mu)$, $Z_{c^2}=Z_{\varepsilon^{-1}} Z_{\mu^{-1}}$, yielding
\begin{equation}
Z_c=1-\frac{2}{3} \frac{e^2}{\varepsilon c^2 v_F} \left[ N_F \left(1-\frac{v^2_F}{c^2} \right) + N_B \frac{c v_F}{v^2_B} \left( \frac{c^2}{v^2_B}-1\right) \right]\: \frac{1}{\epsilon},
\end{equation}
from which we arrive at the flow equations for fine structure constant ($\alpha_F$) [see Eq.~(\ref{alpha-flow})] and velocity of light (c) [see Eq.~(\ref{c-flow})] in 3+1 dimensions.

{\bf Open Access.} This article is distributed under the terms of the Creative Commons Attribution License (CC-BY 4.0), which permits any use, distribution and reproduction in any medium, provided the original author(s) and source are credited.

\end{document}